\RequirePackage{fix-cm}
\documentclass[smallextended]{svjour3}       
\smartqed  
\usepackage{graphicx}
\begin{document}

\title{The conventions for the polarization angle
}


\author{Sperello di Serego Alighieri         
}


\institute{S. di Serego Alighieri \at
              INAF - Osservatorio Astrofisico di Arcetri, Largo E. Fermi 5, 50125 Firenze, Italy \\
              Tel.: +39-055-2752311\\
              \email{sperello@arcetri.astro.it}           
           \and
}

\date{Received: date / Accepted: date}

\maketitle

\begin{abstract}
Since more than a century astronomers measure the position angle of the major axis of the polarization ellipse starting from the North direction and increasing counter-clockwise, when looking at the source. This convention has been enforced by the IAU with a Resolution in 1973. Much later the WMAP satellite, which has observed the polarization of the cosmic microwave background, has unfortunately adopted the opposite convention: the polarization position angle is measured starting from the South and increasing clockwise, when looking at the source. This opposite convention has been followed by most cosmic microwave background polarization experiments and is causing obvious problems and misunderstandings. The attempts and prospects to enforce the official IAU convention are described. 

\end{abstract}

\section{Background}
\label{back}
\begin{figure*}
  \includegraphics[width=0.75\textwidth]{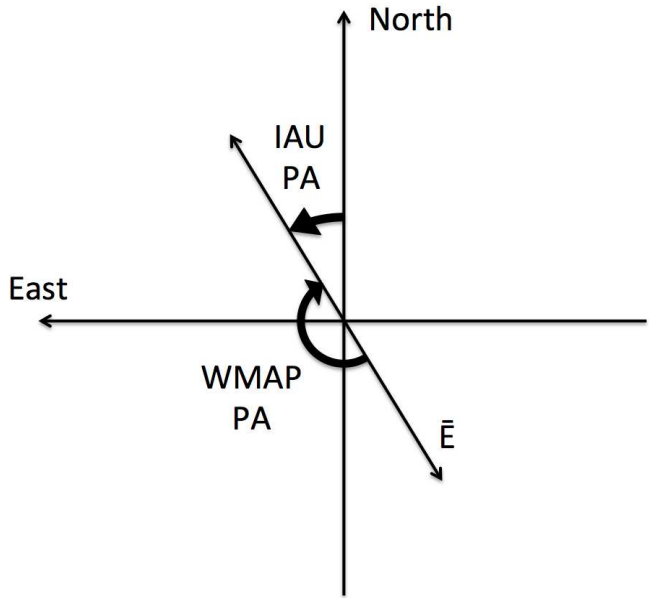}
\caption{The two opposite conventions for the polarization angle. The figure is drawn as when looking at the source and the double arrow represents the plane of vibration of the E-vector.}
\label{fig:1}       
\end{figure*}

Polarimetry of celestial sources is more than two centuries old. Back in 1811 Arago observed the polarization of solar light reflected by the moon. Later the polarization of planets, asteroids, stars and nebulae was measured. These early astronomers made the obvious choice to report the orientation of the polarization as the position angle of the plane of vibration of the E-vector of linear polarization, called the polarization angle (PA) for short. In the more general case of elliptical polarization, PA is the position angle of the major axis of the polarization ellipse. The position angle of a vector in the sky is conventionally measured by astronomers starting from the North direction and going towards East, i.e. increasing counter-clockwise, when looking at the source. Therefore the convention adopted by astronomers for PA is the same they use for the position angle of a vector, with the only difference that the PA, being the position angle of a plane, only goes from 0 to 180 degrees (Fig. \ref{fig:1}). 

This is equivalent to choosing a right-handed frame for which the x-axis corresponds to declination, the y-axis to right ascension and the z-axis to the direction of propagation of the light from the source to us. I should stress that these are merely conventions, since there is nothing physical telling us which way the position angle of a vector or the PA should go in the sky. As for any scientific convention, once it has been set, it is important that all scientists stick to it, in order to avoid confusion and mistakes. Therefore at the IAU General Assembly in Sidney in 1973 Commissions 25 and 40 issued a Resolution enforcing the above convention for PA \cite{IAU74}, in conformity with the IEEE Standard 211 of 1969.

\section{The problem of the two opposite conventions}
\label{problem}
All went well for the PA convention, also when the polarization of the cosmic microwave background (CMB) was discovered in 2002 by DASI \cite{Kov02}, 
since in this paper the IAU convention was used. Also the Cosmic Background Imager \cite{Rea04} used the IAU convention. However the Wilkinson Microwave Anisotropy Probe (WMAP) decided to use the opposite convention (PA increasing clockwise, when looking at the source, see Fig. 1) already in the presentation of their results for the first year of the mission \cite{Kog03}. They did so following a software for the pixelization of data on the sphere \cite{Gor05}; however the paper describing this software does not even mention PA conventions and follows early theoretical definitions of CMB polarization \cite{Zal97}, completely unaware of the convention already followed by all other astronomers. 

This is equivalent to choosing a right-handed frame for which the x-axis corresponds to the opposite of declination (increasing to the South!), the y-axis to the right ascension and the z-axis to the direction of observation from us to the source. This corresponds to a change of the sign of the U Stokes parameter, and is obviously creating misunderstandings and mistakes. 

Since WMAP has been very influential for CMB polarization studies, most CMB polarization experiments, both from the ground and from space, have since followed the ''WMAP convention'' for PA, allegedly to make it easier to confront their results with those of WMAP. This has happened for example for the Planck satellite. In fact Hamaker and Leahy in their report on ''A study of CMB differencing polarimetry with particular reference to Planck'' \cite{Ham04} describe clearly the two opposite PA conventions, say that ''it is a dangerous source of confusion within the Planck project and even more for outside users of Planck data'', and they ''urgently recommend that Planck management exert its influence to have this anomaly corrected before it gets spread more widely''. Unfortunately their recommendation was not followed up timely, the Planck team sticked with the WMAP convention, and most papers with Planck polarization results used it, often without even mentioning it. 

An example of the confusion generated by the opposite WMAP convention is given in the paper \cite{Kau16}, which reviews the measurements of the Cosmic Polarization Rotation, obtained with different methods. In a first version of their paper, still available in {\it https://arxiv.org/abs/1409.8242v1}, the authors had not taken into account that the first measurements with radio galaxies used the IAU convention, while the later ones with the CMB used the WMAP convention. In fact in the first version of their figure 1 measurements taken with the different conventions are mixed together, fortuitously generating the wrong impression that most measurements are negative! The figure was then corrected in the published version of the paper, and in the case of the CPR, whose measurements are consistent with zero, a PA sign change is not dramatic, but confusion is definitely generated. 

When I warned Planck managers about the problem and urged them to adopt the IAU convention, they argued that it would be impossible for them to adapt their software to the IAU convention. However no software change is actually required, just a sign change for the U Stokes parameter in the reduced data. In fact some Planck papers, in particularly those on the polarization of foreground sources, used the IAU convention, by simply inverting the sign of the U Stokes parameter in the Planck data \cite{Ade15}. Therefore the Planck polarization data have been published sometimes using the WMAP convention, and some other time using the IAU convention: a really confusing situation!

\section{Outlook}
\label{out}
I became aware of the double convention problem late in 2014 and since worked towards a solution. The obvious one is that everybody uses the convention set earliest, used by the majority, and officially endorsed by the IAU. 

On December 8, 2015 the General Secretary of the IAU, together with the Presidents of Division B (Facilities, Technologies and Data Science) and Commission B6 (Astronomical Photometry and Polarimetry) have issued a Recommendation that ''all astronomers, including those working on the CMB, follow the IAU Resolution for the the Polarization Angle in all their publications'' (see {\it http://www.iau.org/news/announcements/detail/ann16004/}). This obvious appeal can be adopted easily, independently of any software used for the data. In fact it is easy and straightforward to change the sign of the U Stokes parameter, as clearly demonstrated for example by \cite{Ade15}.

\begin{acknowledgements}
I would like to thank Egidio Landi degli Innocenti for advice and support and the referee for useful suggestions.
\end{acknowledgements}


\end{document}